\documentstyle[12pt,epsfig]{article}

\catcode`\@=11
\long\def\@makefntext#1{
\protect\noindent \hbox to 3.2pt {\hskip-.9pt  
$^{{\ninerm\@thefnmark}}$\hfil}#1\hfill}                

\def\@makefnmark{\hbox to 0pt{$^{\@thefnmark}$\hss}}  
        
\def\ps@myheadings{\let\@mkboth\@gobbletwo
\def\@oddhead{\hbox{}
\rightmark\hfil\ninerm\thepage}   
\def\@oddfoot{}\def\@evenhead{\ninerm\thepage\hfil
\leftmark\hbox{}}\def\@evenfoot{}
\def\sectionmark##1{}\def\subsectionmark##1{}}

\renewcommand{\thefootnote}{\fnsymbol{footnote}}

\newcounter{sectionc}\newcounter{subsectionc}\newcounter{subsubsectionc}
\renewcommand{\section}[1] {\vspace*{0.6cm}\addtocounter{sectionc}{1} 
\setcounter{subsectionc}{0}\setcounter{subsubsectionc}{0}\noindent 
        {\normalsize\bf\thesectionc. #1}\par\vspace*{0.4cm}}
\renewcommand{\subsection}[1] {\vspace*{0.6cm}\addtocounter{subsectionc}{1} 
        \setcounter{subsubsectionc}{0}\noindent 
        {\normalsize\it\thesectionc.\thesubsectionc. #1}\par\vspace*{0.4cm}}
\renewcommand{\subsubsection}[1]
{\vspace*{0.6cm}\addtocounter{subsubsectionc}{1}
        \noindent {\normalsize\rm\thesectionc.\thesubsectionc.\thesubsubsectionc. 
        #1}\par\vspace*{0.4cm}}

\newcounter{appendixc}
\newcounter{subappendixc}[appendixc]
\newcounter{subsubappendixc}[subappendixc]

\renewcommand{\appendix}[1] {\vspace*{0.6cm}
        \refstepcounter{appendixc}
        \setcounter{figure}{0}
        \setcounter{table}{0}
        \setcounter{equation}{0}
        \renewcommand{\thefigure}{\Alph{appendixc}.\arabic{figure}}
        \renewcommand{\thetable}{\Alph{appendixc}.\arabic{table}}
        \renewcommand{\theappendixc}{\Alph{appendixc}}
        \renewcommand{\theequation}{\Alph{appendixc}.\arabic{equation}}
        \noindent{\bf Appendix \theappendixc #1}\par\vspace*{0.4cm}}

\def\abstracts#1{{
        \centering{\begin{minipage}{12.2truecm}\footnotesize\baselineskip=12pt\noindent
        \centerline{\footnotesize ABSTRACT}\vspace*{0.3cm}
        \parindent=0pt #1
        \end{minipage}}\par}} 

\renewenvironment{thebibliography}[1]
        {\begin{list}{\arabic{enumi}.}
        {\usecounter{enumi}\setlength{\parsep}{0pt}
\setlength{\leftmargin 1.25cm}{\rightmargin 0pt}
         \setlength{\itemsep}{0pt} \settowidth
        {\labelwidth}{#1.}\sloppy}}{\end{list}}

\newcounter{itemlistc}
\newcounter{romanlistc}
\newcounter{alphlistc}
\newcounter{arabiclistc}

\newcommand{\fcaption}[1]{
        \refstepcounter{figure}
        \setbox\@tempboxa = \hbox{\footnotesize Fig.~\thefigure. #1}
        \ifdim \wd\@tempboxa > 6in
           {\begin{center}
        \parbox{6in}{\footnotesize\baselineskip=12pt Fig.~\thefigure. #1}
            \end{center}}
        \else
             {\begin{center}
             {\footnotesize Fig.~\thefigure. #1}
              \end{center}}
        \fi}

\newcommand{\tcaption}[1]{
        \refstepcounter{table}
        \setbox\@tempboxa = \hbox{\footnotesize Table~\thetable. #1}
        \ifdim \wd\@tempboxa > 6in
           {\begin{center}
        \parbox{6in}{\footnotesize\baselineskip=12pt Table~\thetable. #1}
            \end{center}}
        \else
             {\begin{center}
             {\footnotesize Table~\thetable. #1}
              \end{center}}
        \fi}

\def\@citex[#1]#2{\if@filesw\immediate\write\@auxout
        {\string\citation{#2}}\fi
\def\@citea{}\@cite{\@for\@citeb:=#2\do
        {\@citea\def\@citea{,}\@ifundefined
        {b@\@citeb}{{\bf ?}\@warning
        {Citation `\@citeb' on page \thepage \space undefined}}
        {\csname b@\@citeb\endcsname}}}{#1}}

\newif\if@cghi
\def\cite{\@cghitrue\@ifnextchar [{\@tempswatrue
        \@citex}{\@tempswafalse\@citex[]}}
\def\citelow{\@cghifalse\@ifnextchar {\@tempswatrue
        \@citex}{\@tempswafalse\@citex[]}}
\def\@cite#1#2{{$\null^{#1}$\if@tempswa\typeout
        {IJCGA warning: optional citation argument 
        ignored: `#2'} \fi}}

\newcommand{\gnufig}[3]{\begin{figure}[t]
\begin{center}
{#3}
\end{center}
\vspace*{0.5cm}
\fcaption{#2}
\label{#1}
\end{figure}}

 1
 1
 1

\font\ninerm=cmr9

\def\be{\begin{equation}}
\def\ee{\end{equation}}
\def\bea{\begin{eqnarray}}
\def\eea{\end{eqnarray}}

\def\qqg{$q {\bar q}g~$}
\def\qq{$q {\bar q}~$}
\def\pp{$p_{\perp}$~}
\newcommand{\gapp}{\begin{array}{c}> \\[-3mm] \sim \end{array} }
\newcommand{\lapp}{\begin{array}{c}< \\[-3mm] \sim \end{array} }

\begin{document}

\centerline{\normalsize\bf Are charm and high-\pp jets the keys}
\baselineskip=16pt
\centerline{\normalsize\bf to understanding diffraction in DIS ?\footnote{Talk given at `New Trends in HERA Physics' Ringberg, May
  1997. To appear in the proceedings. Also available as a preprint, DESY 97-155.}}

\centerline{\footnotesize M.F. McDermott}
\baselineskip=13pt
\centerline{\footnotesize\it DESY, Theory Group, Notkestrasse 85,}
\baselineskip=12pt
\centerline{\footnotesize\it Hamburg, 22607, Germany}
\centerline{\footnotesize E-mail: mcdermot@mail.desy.de}

\vspace*{0.9cm}
\abstracts{Following an introduction which explains some basic 
ideas about diffraction, the semiclassical approach to diffraction in
deep inelastic scattering (DIS) is outlined. 
Some phenomenological tests of this picture, concerning the
$p_{\perp}$ and mass spectra of open charm and jet production are given.}
 
\normalsize\baselineskip=15pt
\setcounter{footnote}{0}
\renewcommand{\thefootnote}{\alph{footnote}}
\section{Introduction}

To introduce the main topic of this talk, I would like, first of all,
to recall some basic ideas concerning diffraction that it is useful to
have at the back of one's mind when thinking about diffraction in high
energy processes such as DIS at small $x$. I will then discuss
predictions for diffractive events containing  either 
high $p_\perp$-jets or charm quarks in the final state,  
for two-gluon exchange models and the semiclassical approach.

\subsection{The optical model}

Let us begin by reminding ourselves of the phenomena of diffraction in
optics. Consider a broad beam of plane polarized light incident on a small
piece of Polaroid, which has its axis of polarization at an  angle,
$\theta$,  to the polarization direction. The component of the light
with its polarization parallel to this axis will be transmitted 
and that perpendicular to it will be absorbed. 
The transmitted wave, just behind the Polaroid now
contains two components, with polarizations parallel and perpendicular
to that of the incident wave. 
A new state, degenerate in energy to the initial one,
has been ``diffracted into existence'' by the piece of Polaroid.  

Good and Walker developed an optical model for diffraction in the high
energy scattering of hadrons as early as 1960~\cite{MFMGW}.
They consider the possibility that a beam of hadrons, of type $A$ and mass
$M$, incident on a heavy nucleus at rest, produces two new particles, $B$
and $C$, of combined rest mass $M^*$, leaving the nucleus in its ground
state. The longitudinal kick given to the nucleus needs to be fairly
small to avoid breaking it up : 
$q_{\|} \le m_{\pi}/A_N^{1/3}$, where $m_{\pi}$ is the mass of the pion
and $A_N$ is the nucleon number of the nucleus. 
As long as hadron $A$ has a three momentum, $P$, of at least 
\be
P_{th} = \frac{(M^2 - M^{*2}) A_N^{1/3}}{m_{\pi}}
\ee
then the process is energetically possible. According to quantum
mechanics the dressed nucleon $|{\tilde A}\rangle$  is a complicated 
superposition of `bare' states, $|B_i\rangle$, which have the same 
quantum numbers as $A$. These, 
in turn, may be re-expanded in terms of a set of `physical' 
dressed states $|\tilde{D_j}\rangle$,
\be
|{\tilde A}\rangle = \sum_i b_{Ai} | B_i\rangle 
= \sum_{ij} b_{Ai} d_{ij} |{\tilde D_j}\rangle \,. 
\ee
For very high beam energies $P \gg P_{th}$, 
the system $A$ spends a very short time in the heavy nucleus
on the timescale of its quantum fluctuations, we may consider 
these fluctuations as essentially frozen during the interaction. 

In the nuclear medium $A$ will have a different expansion, 
$|{\tilde A}\rangle = c_{Ak}|{\tilde C_k}\rangle$. 
Each component $|{\tilde C_k}\rangle$ constitutes an
eigenstate of the interaction and may be absorbed differently 
by the nucleus. For the scattering matrix just behind the target we
have 
\be
|S\rangle= |I\rangle - |T\rangle =  (1-{\bar \eta})|{\tilde A}\rangle + \sum_i ({\bar \eta}-\eta_i) c_{Ai}|C_i\rangle.
\ee
Provided the eigenstates have different absorption coefficients 
$({\bar \eta} \ne \eta_i)$ 
then new physical states $|C_i\rangle$, ``degenerate in mass''
with $A$, may be diffracted into existence by the interaction with the
nuclear target.

\subsection{High energy hadron-hadron diffractive scattering, rapidity gaps,\\ 
light cone co-ordinates and variables for diffractive DIS}

Consider very high energy diffractive scattering of two hadrons:
hadron $A$ of momentum $P_A$, travelling in the $+z$ direction, loses
only a very small amount of its momentum 
and remains intact $(P'_A \gapp 0.99P_A)$; 
hadron $B$, travelling in the opposite direction, breaks up.  
It is very useful to introduce {\it light cone co-ordinates} 
for the patrons in these hadrons:
\bea
q_i^+  & = & q_i^0 + q_i^z =  m_i^{\perp} (\cosh{y_i} + \sinh{y_i}) =
m_i^{\perp} e^{y_i} \,, \label{eq:qplus} \\
q_i^-  & = & q_i^0 - q_i^z =  m_i^{\perp} (\cosh y_i - \sinh y_i) = m_i^{\perp}
e^{-y_i} \,, \\
q_i^{\perp}  & = & (q_i^x, q_i^y) \,, \\
m_i^{\perp} & = & \sqrt{q_i^{\perp 2} + m_i^2} \,, \label{eq:tmass}
\eea
\noindent where $y_i$ is the rapidity of parton $i$ and is approximately
equal to its pseudorapidity $\eta_i$ provided $m_i^2 \ll q_i^{\perp2}$. A
typical parton in hadron $B$ has a large $q^-$ component and a small
$q^+$ component and the converse is true for a parton in hadron $A$.
Momentum conservation
\be
P_A + P_B = P'_A + \sum_{i \epsilon B'} q_i 
\ee
then yields
\be
\xi = \frac{P_A^+ - P_{A}^{'+}}{P_A^+} = \frac{\sum_{i \epsilon B'} q_i^+ - P_B^+}{P_A^+} \approx \frac{\sum_i m_i^{\perp} e^{-(y_A - y_i)}}{m_A} 
\ee
for the momentum fraction, $\xi$, lost by hadron $A$.
We can immediately see that for small $\xi \lapp 0.01$  we will
necessarily have a {\it rapidity gap}, $\eta_{\mbox{gap}} \approx y_A -
y_{i,max}$, in the final state and that we can
get a reasonable experimental measure of $\xi$, even without measuring
the final state hadron, by considering hadrons close to the gap
and those with high $q^{\perp2}_i$ which contribute most in the above
sum (for a more detailed discussion see Appendix A of Collins
{\it et al}~\cite{MFMCHPWW}).

In diffractive DIS one has to integrate over phase space, 
$d^3P'$, of the scattered proton. Integrating over
the angle, one has two additional scattering variables $t$ and
$\xi$. The scattering variables usually used to describe the events
are 
\bea
x = \frac{Q^2}{Q^2+ W^2} & \,,& M^2 = (P-P'+q)^2  \,, \\
t = (P-P')^2  & \,,&  \xi \approx \frac{M^2 + Q^2}{W^2 + Q^2} \,,\\
\beta = \frac{Q^2}{2q.(P-P')} = \frac{Q^2}{M^2 + Q^2 -t} &\approx& 
\frac{Q^2}{M^2 + Q^2} \approx \frac{x}{\xi} \,,
\eea
\noindent where $W^2$ is the centre of mass energy of the 
$\gamma^* P$-subprocess, $q$ and $q^2 = -Q^2$ are the four momentum and
virtuality of the photon, and $M$ is the mass of the diffractive final state. 
In order that the proton does not break up, the momentum 
transfer in the interaction should not be too large: $t \lapp 1$GeV$^2$.
We will be primarily concerned with events for which  $M^2 \approx Q^2 \ll
W^2$, i.e. with the diffractive production of mass states approximately
degenerate with the virtuality of the photon.

\subsection{Eigenstates of diffraction in DIS}

An analogous process to the optical model discussed above is the production of
an $e^+ e^-$ pair from a highly energetic virtual photon in a Coulomb field,
such as that surrounding a heavy nucleus. The pair is a virtual 
fluctuation of the virtual photon and, provided the
individual virtualities of the fermions are small, will require only a
small `kick' from the field to push them onto mass shell. 

This notion forms the basis of the semiclassical approach to 
diffraction in DIS. Viewed from the proton's 
rest frame, DIS at very high energies (small $x$) 
corresponds to the scattering 
of `frozen' parton configurations in the virtual photon from the colour 
field of the proton. Those configurations in which one of the partons 
carries only a small fraction of the photon's longitudinal momentum
and has a low transverse momentum have a 
small `off-shellness' and may be diffracted into existence by receiving a 
small momentum transfer from the proton. The presence of this 
`wee' parton in the photon implies that the  fluctuation develops a 
large transverse size by the time it reaches the proton. 

To see this, consider the lowest order fluctuation of the photon
into a \qq-pair, of transverse momentum 
$p^2_\perp$. Momentum conservation at the photon vertex gives
\be
\frac{\Delta^2_p}{1-\alpha} + \frac{\Delta^2_l}{\alpha} = - \left(
Q^2 + \frac{p^2_\perp + m_q^2}{\alpha(1-\alpha)} \right),
\label{eq:flucqq}
\ee
\noindent where $l, \Delta^2_l$ and $m_q$  are the four-momentum,
virtuality and mass of the anti-quark;  $\alpha = l_z/q_z \approx
l_0/q_0$ is longitudinal momentum fraction of the photon carried by
the anti-quark; $\Delta^2_p$ is the quark virtuality. 
Hence, in order to produce partons that are close to mass shell we
need the asymmetric configurations in which $\alpha$ or $1-\alpha \ll 1$,
$ p^2_\perp \ll Q^2, m_q^2 = 0$. 
The uncertainty principle then gives the following
estimate for the lifetime of these fluctuations
\be
\Delta \tau \sim \frac{1}{\Delta E} \sim \frac{1}{m_p x_{bj}} \left(
\frac{Q^2}{Q^2 + p^2_\perp/\alpha(1-\alpha)} \right)\,,
\ee
\noindent which is much longer, at small $x_{bj}$, than the typical 
timescales involved in the interaction of the partons with the proton, 
which are of the order of the inverse of the mass, $m_p$, of the proton.

Miettinen and Pumplin~\cite{MFMMP} were the first to recognize that 
frozen parton fluctuations, in general, correspond to the eigenstates of
diffraction. A very nice discussion of many of the points
made here can be found in Chapter 7 of the recent text book by 
Forshaw and Ross~\cite{MFMFR}.

\section{Two-gluon exchange models}

The simplest QCD model one can think of for diffraction is the
exchange of two gluons in a colour singlet in the $t$-channel.
Several recent papers present predictions for diffractive production
of charm~\cite{MFMTWOGC} and high $p_{\perp}$-jets~\cite{MFMTWOGPT},
for the pure $q \bar q$ final state,  
which have this mechanism of diffractive exchange in common. 
Unfortunately a complete ${\cal O}(\alpha_s)$ calculation, 
which would also include a gluon in the final state is not yet
available. In terms of the diffractive structure function, these 
corrections are expected to be important when the diffractive mass is
large compared to $Q^2$. The differential cross sections for 
transversely and longitudinally  polarized photons can be written, 
in the double leading logarithmic
approximation, in terms of the square of the gluon density, $G(\xi)$,
as follows 
\bea
\frac{d^2 \sigma_L}{d \alpha dp_{\perp}^2} & = & \frac{2\sum_q e_q^2\alpha_{em}\alpha_s^2
\pi^2[\xi G(\xi, (p^2_\perp + m_q^2)/(1-\beta))]^2C}{3(a^2+p_\perp^2)^6}[\alpha(1-\alpha)]^2Q^2
(a^2-p_\perp^2)^2  \,,  \nonumber \\
& & \label{eq:twogqql} \\
\frac{d^2 \sigma_T}{d\alpha dp_{\perp}^2} & = & \frac{\sum_q e_q^2\alpha_{em}\alpha_s^2\pi^2
[\xi G(\xi, (p^2_\perp + m_q^2)/(1-\beta))]^2C}{6(a^2+p_\perp^2)^6} \nonumber \\
& \times & \left[4(\alpha^2+(1-\alpha)^2)p_\perp^2
a^4 + m_q^2(a^2-p_\perp^2)^2\right]\,, \label{eq:twogqqt} \\
a^2 & = & \alpha (1-\alpha) Q^2 + m_q^2 \,. 
\eea
\noindent The factor $C$ parameterizes the required extrapolation from 
$t\approx 0$ to the integrated cross section, $C \approx \Lambda^2$
(where $\Lambda$ is a typical hadronic scale).

\section{The semiclassical approach}

In this section predictions for the diffractive  production of heavy
flavours and high-$p_{\perp}$ jets are presented for \qq and \qqg
final states, in the semiclassical approach~\cite{MFMBHM}. 
Phenomenological implications of this approach have been investigated in recent
preprints~\cite{MFMBHMC}~\cite{MFMBHMPT}.

\subsection{\qq final states}

The longitudinal and transverse differential cross sections for the leading 
order $q \bar{q}$ ~fluctuation are as follows
\begin{eqnarray}
\frac{d\sigma_L}{d\alpha dp'^{2}_\perp}&=&\frac{ 4 \Sigma_q e_q^2
\alpha_{em}}{3(2\pi)} [\alpha(1-\alpha)]^2 \int_{x_\perp}\left|\int\frac{d^2p_\perp}
{(2\pi)^2} \frac{\mbox{tr}\tilde{W}^{\cal F}_{x_\perp}(p_\perp'\!-\!p_\perp)}
{a^2+p_\perp^2}\right|^2\,, \label{eq:slqq} \\
\frac{d\sigma_T}{d\alpha dp'^{2}_\perp}&=&\frac{\Sigma_q e_q^2 \alpha_{em}}{3 (2\pi)} 
(\alpha^2+(1-\alpha)^2) f^{\cal F}(a^2,p_\perp')\,, \label{eq:stqq} \\ 
f^{\cal F}(a^2,p_\perp') & = &\int_{x_\perp}\left|\int\frac{d^2p_\perp}
{(2\pi)^2} \frac{ p_\perp \mbox{tr}\tilde{W}^{\cal F}_{x_\perp}(p_\perp'\!-\!p_\perp)}
{a^2+p_\perp^2}\right|^2\,, \label{eq:ff} \\
a^2 & = & \alpha (1- \alpha) Q^2 + m_q^2 \,.
\end{eqnarray}

\noindent The eikonal factor 
\be
W^{\cal F}_{x_\perp}(y_\perp)=U^\dagger(x_\perp+y_\perp)U(x_\perp)-1\,\label{wa}
\ee
is built from the non-Abelian eikonal factors $U$ and $U^\dagger$ of
the quark 
and antiquark whose light-like paths penetrate the colour field of the 
proton at transverse positions $x_\perp$ and $x_\perp+y_\perp$,
respectively. The superscript ${\cal F}$ is used because the quarks
are in the fundamental representation of the gauge group. 
The function $f^{\cal F}$ depends
on the Fourier transform of the colour trace of the eikonal factor
which depends in general on the transverse momentum lost by parton. 
Taking its trace projects onto the colour singlet
configurations relevant for diffraction.
For a soft colour field, large transverse momentum transfers 
are exponentially suppressed.  The factors in the
denominator and numerator in the integrand of Eq.(\ref{eq:ff}) reflect the
propagator and spin of this parton, respectively.

In the region $\alpha(1-\alpha)Q^2 > \Lambda^2$,
we can expand the denominator of Eq.(\ref{eq:ff}) in the momentum
transfer lost by the quark and, from the properties of the eikonal
function, show that there is no leading twist diffraction in this region. 
For $\alpha(1-\alpha)Q^2 < \Lambda^2$, for low $p_{\perp}$ quarks, 
i.e. the asymmetric configurations, 
this expansion is no longer possible, at least for
massless quarks, since the denominator becomes small. 
As a result one gets a leading twist 
contribution to the transverse cross section in this region. 
The addition powers of $\alpha(1-\alpha)$ in Eq.(\ref{eq:slqq})
ensure that the contribution to the longitudinal cross section is
higher twist.

In the case of a massive fermion, such as the charm quark, the
expansion is always possible since the denominator is limited by the
fermion mass and the cross section for diffractive charm production is
suppressed by $\Lambda^2/m_c^2$. Similarly high-$p_{\perp}$
configurations are suppressed by $\Lambda^2/p^2_{\perp}$. All of this can
be seen indirectly from size of the virtualities in Eq.(\ref{eq:flucqq}).

The integration over $t$ is implicit in Eq.(\ref{eq:ff}), as reflected
in the single integration over $x_{\perp}$. Reinstating the
$t$-integration and expanding in $y_\perp$ (for high-$p_{\perp}$
configurations) we can relate the resulting $x_{\perp}$-integral 
to the gluon density and the cross section, $\sigma(y_{\perp})$, 
for scattering a small dipole of size $y_{\perp}$ from the proton,
\be
\int_{x_\perp} \mbox{tr} W_{x_{\perp}} (y_{\perp}) = \frac{3 \sigma (y^2_{\perp})}{2}
= \frac{\pi^2}{2} \alpha_s y^2_{\perp} \xi G(\xi)\,.
\ee
The differential cross sections are then exactly those given in
Eqs.(\ref{eq:twogqql},\ref{eq:twogqqt}) for the two-gluon exchange models. 
We see that the mechanism for the diffractive production of 
jets or open charm is essentially `hard' and corresponds to the
exchange of two gluons in the $t$-channel.

\subsection{\qqg final states}

\gnufig{F:MFM:QQG}{Dominant partonic process for diffractive
charm and high-$p_{\perp}$ jet production in the semiclassical
approach.}
{\hspace{0.7cm}
\epsfig{file=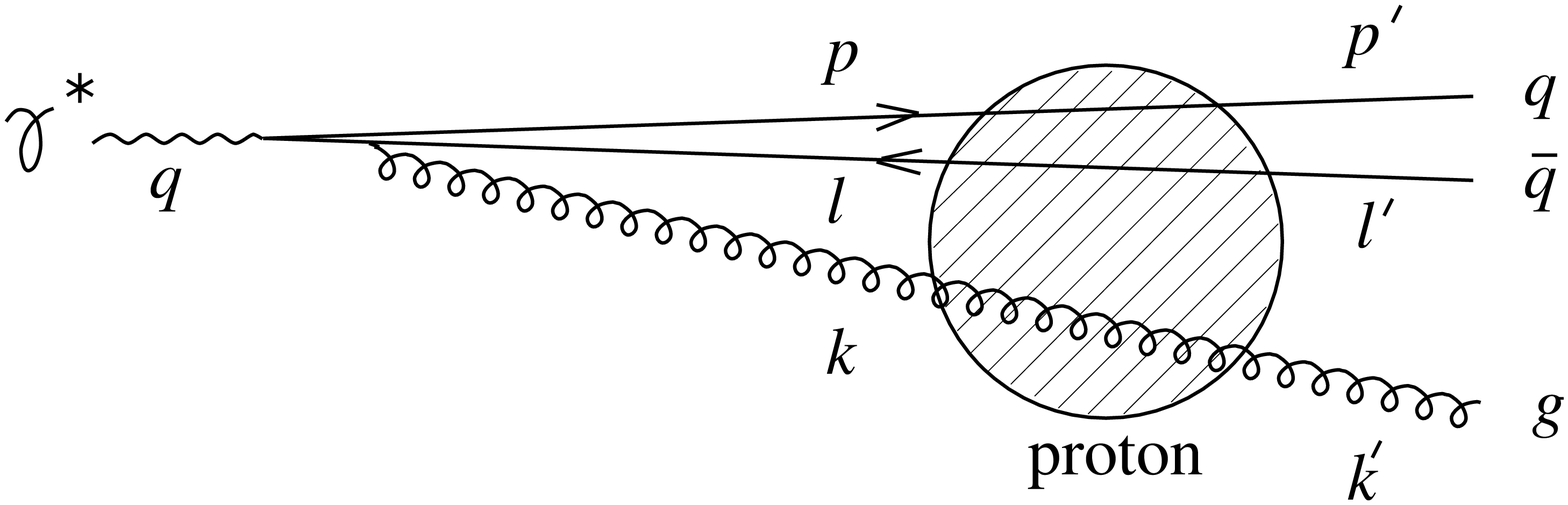,width=12cm}}

The ${\cal O}(\alpha_s)$ calculation with an additional gluon in the
photon fluctuation proceeds along similar conceptual lines. 
Momentum conservation, for the fluctuation with a wee gluon
and two high $p_\perp$-jets, now gives
\be
\frac{\Delta^2_p}{1-\alpha} + \frac{\Delta^2_l}{\alpha} +
\frac{\Delta^2_k}{\alpha'} 
= - \left(
Q^2 + \frac{p^2_\perp + m_q^2}{\alpha(1-\alpha)} +
\frac{k^2_{\perp}}{\alpha'}  \right),
\label{eq:flucqqg}
\ee
for a gluon of transverse momentum $k_{\perp}$, virtuality $
\Delta^2_k $ and momentum fraction $\alpha'$. Leading twist
diffraction can now result in two cases: either the gluon is wee 
($\alpha' \ll 1, k^2_{\perp} \ll Q^2 $) or the fermion is wee and
massless (to avoid the mass suppression). The other two partons may
emerge in a high transverse momentum configuration to produce jets.
For the case with a wee gluon, shown in Fig.(\ref{F:MFM:QQG}), we have 
\bea
\frac{d\sigma_L}{d\alpha dp_\perp^2d\alpha'dk_\perp'^2}&
=&\frac{\sum_q e_q^2\alpha_{em}\alpha_s}{16\pi^2}\,\frac{\alpha'Q^2
p_\perp^2}{[\alpha(1\!-\!\alpha)]^2N^4}f^{\cal A}(\alpha'N^2,k_\perp')
\,, \label{eq:qqgl} \\ 
\frac{d\sigma_T}{d\alpha dp_\perp^2d\alpha'dk_\perp'^2}& =&
\frac{\sum_q e_q^2\alpha_{em}\alpha_s}{128\pi^2}\,\frac{\alpha'
\left\{[\alpha^2+(1\!-\!\alpha)^2]\,[p_\perp^4+a^4]+2p_\perp^2m_q^2\right\}}
{[\alpha(1\!-\!\alpha)]^4N^4} f^{\cal A}(\alpha'N^2,k_\perp')\,,  \nonumber \\
& & \label{eq:qqgt} \\
f^{\cal A}(\alpha'N^2,k_\perp') &=& \int_{x_\perp}\left|\int\frac{d^2k_\perp}
{(2\pi)^2}\left(\delta^{ij}+\frac{2k_\perp^ik_\perp^j}{\alpha'N^2}\right)
\frac{\mbox{tr}\tilde{W}^{\cal A}_{x_\perp}(k_\perp'\!-\!k_\perp)}
{\alpha'N^2+k_\perp^2}\right|^2 \,, \label{eq:fa}  \\
N^2&=&Q^2+\frac{p_\perp^2+m_q^2}{\alpha(1\!-\!\alpha)} \,.
\eea

A heavy quark mass or high-$p_{\perp}$ ensures that the $q {\bar q}$ 
pair stays small in transverse space and behaves like a gluon terms of
colour. The superscript ${\cal A}$ is used 
since we are effectively testing the proton's  field with two ``gluons''. 
The factor in round brackets and the denominator factor in
Eq.(\ref{eq:fa}) reflect the spin and propagator of the wee gluon, 
respectively.
Integration over the final state variables
of the wee gluon in the leading $\ln(1/x)$ approximation gives
\bea
\frac{d\sigma_L}{d\alpha dp_\perp^2} & = &\frac{\sum_q e_q^2\alpha_{em}\alpha_s
\ln(1/x)h_{\cal A}}{2\pi^3(a^2+p_\perp^2)^4}\,[\alpha(1-\alpha)]^2Q^2 
p_\perp^2 \,, \label{eq:sleg} \\
\frac{d\sigma_T}{d\alpha dp_\perp^2} & = &\frac{\sum_q e_q^2\alpha_{em}\alpha_s
\ln(1/x)h_{\cal A}}{16\pi^3(a^2+p_\perp^2)^4}\,\left[(\alpha^2+
(1\!-\!\alpha)^2)\,(p_\perp^4+a^4)+2p_\perp^2m_q^2\right]\,, \label{eq:steg} \\
h_{\cal A} & = & \int_{y_\perp}\int_{x_\perp}\frac{\left|\mbox{tr}
W^{\cal A}_{x_\perp}(y_\perp)\right|^2}{y_\perp^4} \,. 
\eea

In the wee massless fermion case we have

\begin{eqnarray}
\frac{d\sigma_L}{d\alpha dp_\perp^2d\alpha'dk_\perp'^2}&
=&\frac{4 \Sigma_q e_q^2  \alpha_{em}\alpha_s}{9\pi^2}\,\frac{Q^2}
{[\alpha(1\!-\!\alpha)]^2N^4}f^{\cal F}(\alpha'N^2,k_\perp')\,,\label{eq:sleq}
\\ \nonumber\\
\frac{d\sigma_T}{d\alpha dp_\perp^2d\alpha'dk_\perp'^2}&
=&\frac{\Sigma_q e_q^2 \alpha_{em}\alpha_s}{9\pi^2 p^2_\perp N^4}\,
\left[ N^4 - 2Q^2(N^2+Q^2) +  \frac{N^4+Q^4}{\alpha(1\!-\!\alpha)} \right]
 f^{\cal F}(\alpha'N^2,k_\perp')\,, \nonumber \\
& & \label{eq:steq} 
\eea
\noindent where the functional $f^{\cal F}$ is given in Eq.(\ref{eq:ff}).
Now $\alpha',k_\perp'$ refer to the final state variables of
the slow quark (anti-quark) 
and $\alpha, p_\perp$ to the fast anti-quark (quark). 
Since the fast anti-quark and gluon stay very 
close together in transverse space, in terms of colour they behave like
an anti-quark.  Integration gives 
\begin{eqnarray}
\frac{d\sigma_L}{d\alpha dp_\perp^2}&=&\frac{16 \Sigma_q e_q^2\alpha_{em}\alpha_s
}{27\pi^3} \frac{Q^2}{[\alpha(1-\alpha)] N^6}\, h_{\cal F} \,, \label{eq:slq}
\\ \nonumber\\
\frac{d\sigma_T}{d\alpha dp_\perp^2}&=&\frac{4 \Sigma_q
  e_q^2\alpha_{em}\alpha_s}{27 \pi^3 N^6 p^2_\perp} 
\left[ N^4 - 2Q^2(N^2+Q^2) +
  \frac{(N^4+Q^4)}{\alpha(1\!-\!\alpha)} \right]
\,h_{\cal F}\,, \label{eq:stq}
\end{eqnarray}

\noindent with $h_{\cal F}$ defined like $h_{\cal A}$.

Comparing Eqs.(\ref{eq:sleg},\ref{eq:steg}) with
Eqs.(\ref{eq:slq},\ref{eq:stq}) we see that the 
configurations with a wee gluon are enhanced by
$\ln(1/x)$ at small $x$ relative to those with a wee fermion. 
In addition there may be an additional 
large suppression of the slow fermion contributions due to colour 
factors ($h_{\cal A} \approx 16 h_{\cal F}$ \cite{MFMBHMC}). As a
result we claim 
that the configurations with a wee gluon dominate over those with a
wee fermion in the small-$x$ region relevant to diffraction and we shall
ignore the latter from now on. In contrast to the $q {\bar q}$ case
the diffractive production of high-$p_{\perp}$ jets and charm is now
`soft' and relies on the non-perturbative interaction of the wee gluon
with the proton.

\subsection{Boost to the Breit frame}

We now consider what happens when one boosts from the Proton's rest
frame to the Breit frame (in which q = (0,0,0,Q) and the proton is fast). 
As an example we will consider the $q {\bar q} g$
fluctuation with a wee gluon. The general relations for the boost
along the $z$-axis, in light cone co-ordinates of Eqs.(\ref{eq:qplus}-\ref{eq:tmass}),
are 
\be
a'^{+} = \gamma (1-\beta)a^+ ~~~;~~~ a'^{-} = \gamma (1+\beta)a^- \,.
\ee
\noindent For this particular boost
\be
\gamma (1-\beta) = \frac{m_p x}{|Q|} = \frac{1}{\gamma (1+\beta)}\,. 
\ee
The typical four momentum of the slow gluon before interaction is 
\be
k = (\frac{\Lambda^2}{m_p x}, - \Lambda x, {\vec k}^{\perp}) \,, 
\ee 
where, crucially, its `-' component is {\it negative}. The boost merely
rescales the components to give
\be
k' = (\frac{\Lambda^2}{|Q|}, - \frac{\Lambda |Q|}{m_p}, {\vec k}^{\perp})\,.
\ee 
\noindent This negative energy outgoing gluon may then be reinterpreted as a
positive energy {\it incoming} parton in the proton~\cite{MFMAH}. 
In contrast, the final state wee gluon is forced to have a positive
`-' component because it is on shell and ends up in the final state.
So, in the Breit frame, the process of Fig.(\ref{F:MFM:QQG}) may be
interpreted as boson-gluon fusion with an additional gluon in the final state.

\section{Features of diffractive final states}

We propose two phenomenological tests which reveal the nature of the underlying
process and reflect the mechanism of colour neutralization which
produces the different diffractive final states. Firstly, one may ask
how many high-$p_{\perp}$ events survive above a given minimum
$p_{\perp}^2$. To examine this we plot the quantity 
\bea
\sigma(p^2_{\perp,\mbox{\footnotesize min}}) &=& 
\int_{p^2_{\perp,\mbox{\scriptsize min}}}^{\infty}
{dp^2_\perp} \int_{0}^{1} d\alpha \frac{d^2 \sigma}{dp_\perp^2
  d\alpha} 
\eea
\noindent in Fig.(\ref{F:MFM:PSQ}) for massless quarks. Each curve is
normalized to its value at $p^2_{\perp,\mbox{\footnotesize min}}= 5$~GeV$^2$.
The figure clearly reveals a much harder spectrum for the \qqg final states than
for $q {\bar q}$ . For diffractive charm production we may 
set  ${p^2_{\perp,\mbox{\scriptsize min}}} = 0$ and get similar results~\cite{MFMBHMC}.

\gnufig{F:MFM:PSQ}{The fraction of diffractive events with $p'^2_{\perp}$ above 
$p'^2_{\perp,\mbox{\footnotesize min}}$ for $Q^2$ of 10~GeV$^2$ and~100
GeV$^2$ (lower and upper curve in each pair).}{
\setlength{\unitlength}{0.1bp}
\begin{picture}(4320,2592)(0,0)
\includegraphics{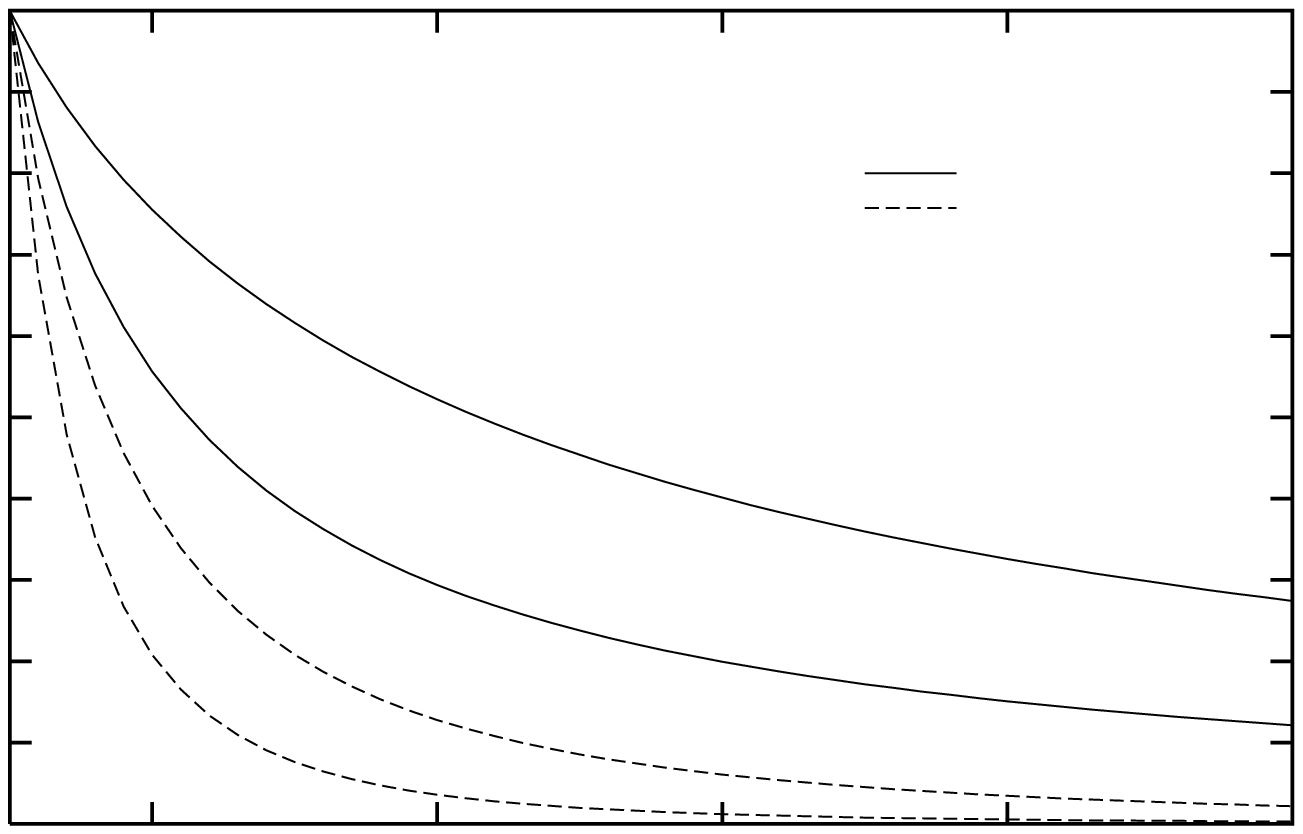}
\put(2876,1924){\makebox(0,0)[r]{ }}
\put(2876,2024){\makebox(0,0)[r]{ }}
\put(2679,1907){\makebox(0,0)[l]{$q \bar{q} ~~ $}}
\put(2679,2024){\makebox(0,0)[l]{$q \bar{q} g $}}
\put(2023,-118){\makebox(0,0)[l]{$p'^2_{\perp,\mbox{min}} ~(\mbox{GeV}^2)$}}
\put(100,1321){%
\makebox(0,0)[b]{\shortstack{Fraction  of Events}}%
}
\put(4157,50){\makebox(0,0){50}}
\put(3336,50){\makebox(0,0){40}}
\put(2515,50){\makebox(0,0){30}}
\put(1694,50){\makebox(0,0){20}}
\put(873,50){\makebox(0,0){10}}
\put(413,2492){\makebox(0,0)[r]{1}}
\put(413,2258){\makebox(0,0)[r]{0.9}}
\put(413,2024){\makebox(0,0)[r]{0.8}}
\put(413,1789){\makebox(0,0)[r]{0.7}}
\put(413,1555){\makebox(0,0)[r]{0.6}}
\put(413,1321){\makebox(0,0)[r]{0.5}}
\put(413,1087){\makebox(0,0)[r]{0.4}}
\put(413,853){\makebox(0,0)[r]{0.3}}
\put(413,618){\makebox(0,0)[r]{0.2}}
\put(413,384){\makebox(0,0)[r]{0.1}}
\put(413,150){\makebox(0,0)[r]{0}}
\end{picture}}

\gnufig{F:SP}{Distributions in $M^2$ and $M_j^2$ of diffractive events 
originating from \qq\ and \qqg\ final states.}{
\setlength{\unitlength}{0.1bp}
\begin{picture}(3600,2160)(0,0)
\includegraphics{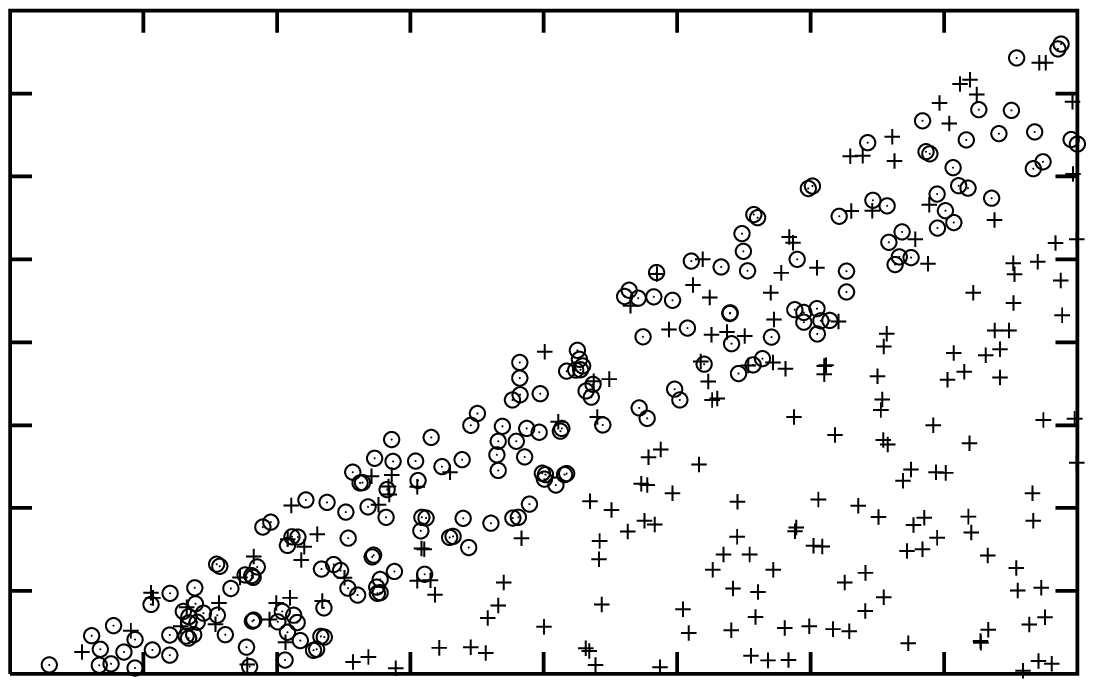}
\put(1132,1702){\makebox(0,0)[l]{$+ ~- ~q \bar{q} g$}}
\put(1132,1821){\makebox(0,0)[l]{$\odot ~- ~q \bar{q}  $}}
\put(1631,-136){\makebox(0,0)[l]{$M^2 ~(\mbox{GeV}^2)$}}
\put(-481,1105){\makebox(0,0)[l]{$M^2_{j} ~(\mbox{GeV}^2)$ }}
\put(3437,50){\makebox(0,0){100}}
\put(3053,50){\makebox(0,0){90}}
\put(2669,50){\makebox(0,0){80}}
\put(2284,50){\makebox(0,0){70}}
\put(1900,50){\makebox(0,0){60}}
\put(1516,50){\makebox(0,0){50}}
\put(1132,50){\makebox(0,0){40}}
\put(747,50){\makebox(0,0){30}}
\put(363,50){\makebox(0,0){20}}
\put(313,2060){\makebox(0,0)[r]{100}}
\put(313,1821){\makebox(0,0)[r]{90}}
\put(313,1583){\makebox(0,0)[r]{80}}
\put(313,1344){\makebox(0,0)[r]{70}}
\put(313,1105){\makebox(0,0)[r]{60}}
\put(313,866){\makebox(0,0)[r]{50}}
\put(313,628){\makebox(0,0)[r]{40}}
\put(313,389){\makebox(0,0)[r]{30}}
\put(313,150){\makebox(0,0)[r]{20}}
\end{picture}}

\gnufig{F:MFM:MSQ}{Normalized mass spectra for the 
$c {\bar c} g$ and $c {\bar c}$
final states calculated in the semiclassical approach}{
\vspace*{-0.7cm}
\hspace*{1.4cm}
\setlength{\unitlength}{0.1bp}
\begin{picture}(3600,2160)(0,0)
\includegraphics{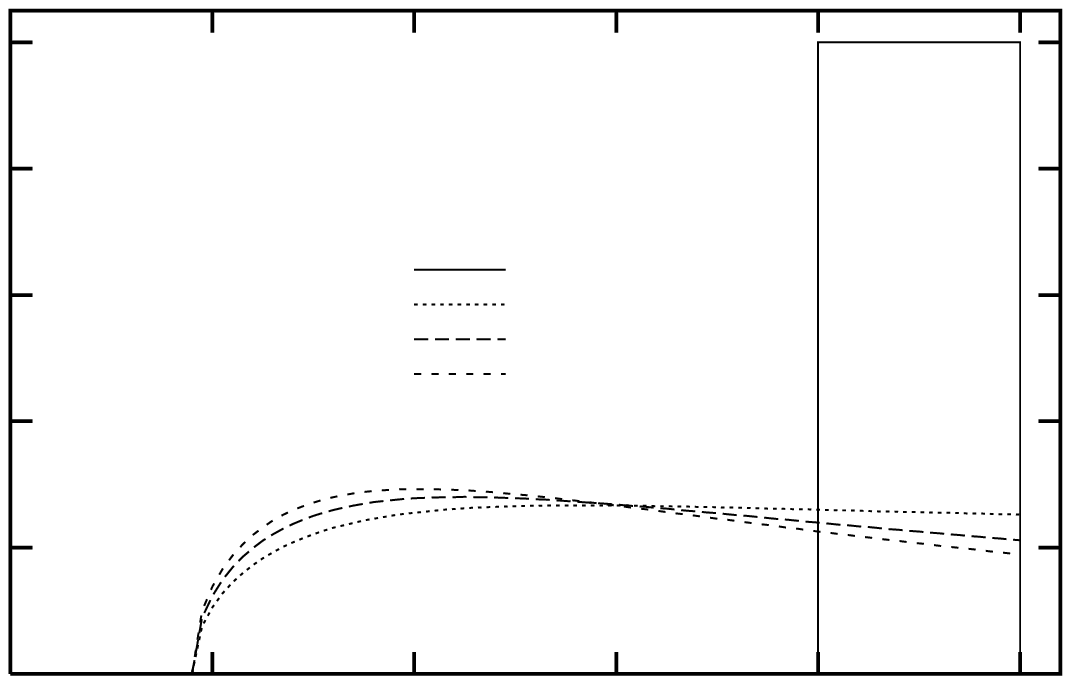}
\put(1526,1014){\makebox(0,0)[r]{$C_s = 2.0$}}
\put(1526,1114){\makebox(0,0)[r]{$C_s = 1.0$}}
\put(1526,1214){\makebox(0,0)[r]{$C_s = 0.5$}}
\put(1526,1314){\makebox(0,0)[r]{$c {\bar c}$ final state}}
\put(1029,1551){\makebox(0,0)[l]{$Q^2 ~= 50 \mbox{~GeV}^2$}}
\put(1751,-70){\makebox(0,0)[l]{$M^2_{c{\bar c}} ~(\mbox{GeV}^2)$}}
\put(-400,1096){\makebox(0,0)[l]{{\Large $\frac{d^2 \sigma_{{\tiny T}}}
{dM^2 dM^2_{c{\bar c}} } $}}}
\put(1038,1442){\makebox(0,0)[l]{$M^2 = 50 \mbox{~GeV}^2$}}
\put(3321,50){\makebox(0,0){50}}
\put(2739,50){\makebox(0,0){40}}
\put(2158,50){\makebox(0,0){30}}
\put(1576,50){\makebox(0,0){20}}
\put(995,50){\makebox(0,0){10}}
\put(413,50){\makebox(0,0){0}}
\put(363,1969){\makebox(0,0)[r]{0.1}}
\put(363,1605){\makebox(0,0)[r]{0.08}}
\put(363,1241){\makebox(0,0)[r]{0.06}}
\put(363,878){\makebox(0,0)[r]{0.04}}
\put(363,514){\makebox(0,0)[r]{0.02}}
\put(363,150){\makebox(0,0)[r]{0}}
\vspace{1.0cm}
\end{picture}}
We may also examine the expected mass spectra for the different final states.
Let $M_j$ be the invariant mass of the two-jet system in diffractive
events.  The  measurement of this observable provides, in principle, a clean distinction 
between \qq\ final states, where $M_j^2=M^2$, and \qqg\ final states, where 
$M_j^2<M^2$. In practice, however, this requires the contribution of the 
wee gluon to the diffractive mass, which is responsible for the difference 
between $M^2$ and $M_j^2$, to be sufficiently large. To quantify the 
expectation within the semiclassical approach we consider the transverse 
photon contribution to the differential diffractive cross section 
$d\sigma/dM^2dM_j^2$.

To illustrate the experimental implications for high-$p_{\perp}$ jets
we have shown in Fig.~(\ref{F:SP}) the positions in $M^2$ and 
$M_j^2$ of two sets of 200 events, scattered randomly according to the two 
above distributions. The $\delta$-function of the $q {\bar q}$ case has been 
replaced by a uniform distribution of $M_j^2$ in a band $M^2>M_j^2>M^2-20$ 
GeV$^2$. This  allows for hadronization effects and, more importantly, for 
a large experimental uncertainty of the mass of the jet system. The
scatter plot clearly exhibits the distinctive features of the
underlying partonic processes, even for this limited number of events. With sufficient 
statistics a determination of the relative weight of soft colour-singlet 
exchange, relevant for  \qqg final states, and hard colour-singlet exchange, 
relevant for \qq final states, should be feasible.

Clearly this is also possible for diffractive charm.
In Fig.(\ref{F:MFM:MSQ}) normalized mass spectra comparing the mass of
the charm pair, $M_{c{\bar c}}$, with the total diffractive mass, $M$,
are shown.  Here the pure $c {\bar c}$ final state  $M_{c{\bar c}} = M$,
is represented by a block of width $10 $~GeV$^2$. The other curves in 
the figure represent the $c{\bar c}g$ final state of Fig.(\ref{F:MFM:QQG})
Details about how the latter curves were arrived at and the meaning 
of $C_s$ are given in the preprint~\cite{MFMBHMC}.

First results on open charm production in diffraction for the 1994
running period were presented by the H1 collab., at Warsaw~\cite{MFMH1C}. 
Both H1 and ZEUS have recently presented results on open charm and
high-$p_{\perp}$ jets in diffractive DIS~{\cite{MFMCHI}}.
Given the increase in statistics of HERA for the 1996 running period, it
is hoped that these phenomenological tests may be performed very soon.

An additional means of distinguishing the underlying diffractive
mechanism is the energy dependence of the two processes. In the two
gluon model a steeply rising gluon density, taken from a fit to $F_2$
for example, produces a rise in diffractive charm events that 
is twice as steep with energy. In contrast the semiclassical approach
the  energy dependence is flat, at least at this order, 
corresponding to a classical bremstrahl spectrum of gluons.


\section{References}

\begin{thebibliography}{100}

\bibitem{MFMGW}
M. L. Good and W. D. Walker {\it Phys. Rev.} {\bf 120} (1960) 1857, 1860. 

\bibitem{MFMCHPWW}
J. Collins {\it et al}, {\it Phys. Rev.} {\bf D51} (1995) 3182.

\bibitem{MFMMP}
H. I. Miettinen and J. Pumplin {\it Phys. Rev.} {\bf D18} (1978) 1696.

\bibitem{MFMFR} J. R. Forshaw and D. A.  Ross, {\it
Quantum Chromodynamics and the Pomeron}, (CUP, Cambridge, 1997).


\bibitem{MFMTWOGC}
E. M. Levin, A. D. Martin, M. G. Ryskin, and T. Teubner, hep-ph/9606443;\\
M. Genovese, N. N. Nikolaev, B. G. Zakharov, {\it Phys. Lett.} {\bf
B378} (1996) 347; \\
H. Lotter, preprint, DESY 96-260, hep-ph/9612415;\\
M. Diehl, preprint, CPTH-S492-0197, hep-ph/9701252.

\bibitem{MFMTWOGPT}   
N. N. Nikolaev and B. G. Zakharov, {\it J. Exp. Theor. Phys.} {\bf 78}  (1994) 598;\\
M. Diehl, {\it Z. Phys.} {\bf C66} (1995) 181;
J. Bartels, H. Lotter and M. W\"usthoff, {\it Phys. Lett.} {\bf B379} 
(1996) 239;
J. Bartels and M. W\"usthoff, {\it J. Phys.} {\bf G22} (1996) 929;
M. W\"usthoff, preprint, ANL-HEP-PR-97-03, hep-ph/9702201.

\bibitem{MFMBHM}
W. Buchm\"{u}ller and A. Hebecker, {\it Nucl. Phys.} {\bf B476} (1996) 203;\\
W. Buchm\"{u}ller, M.~F. McDermott and A. Hebecker, \\
{\it Nucl. Phys.} {\bf B487} (1997) 283; {\it erratum ibid}.

\bibitem{MFMBHMC}
W. Buchm\"uller, M. F. McDermott, and A. Hebecker, \\
{\it Phys. Lett.} {\bf B404} (1997) 353.

\bibitem{MFMBHMPT}
W. Buchm\"uller, M.F. McDermott, and A. Hebecker, hep-ph/9706354.

\bibitem{MFMAH}
A. Hebecker, preprint DAMTP-97-10, hep-ph/9702373.

\bibitem{MFMH1C}
J Phillips, H1. Collab., {\it Proc. of XXVIII ICHEP} (Warsaw, 1996), 
Vol 1, 623.

\bibitem{MFMCHI}
C. Cormack, H1. Collab., T. Terron, ZEUS Collab., Talks at DIS 97,
Chicago, Il., April 1997.

\end{thebibliography}
\end{document}